\documentclass[11pt]{article}
\usepackage[utf8]{inputenc}
\usepackage{amssymb,amsmath}
\usepackage{bm} 
\usepackage{booktabs} 
\usepackage{array}
\usepackage{latexsym}
\usepackage{graphicx}
\usepackage{color}
\usepackage{datetime}
\usepackage[nosort]{cite}
\usepackage{verbatim}
\usepackage{enumerate}
\usepackage{chngpage} 
\usepackage{mathrsfs}
\usepackage{euscript}
\usepackage{psfrag}

\usepackage{datetime}

\usepackage[nosort]{cite}
\usepackage{chngpage} 
\usepackage{setspace}
\usepackage{tensor}
\usepackage{physics}
\usepackage{tensor}

\usepackage{mciteplus}

\usepackage[colorlinks=true,      linkcolor=blue,      urlcolor=blue,      
filecolor=blue,      citecolor=blue,       pdfstartview=FitH,     
pdfpagemode=UseNone,      bookmarksopen=true]{hyperref}  
\usepackage[all]{hypcap}     

\definecolor{cardinal}{rgb}{0.6,0,0}
\definecolor{darkgreen}{rgb}{0,0.4,0}
\definecolor{golden}{rgb}{0.92, 0.7, 0}
\definecolor{midnight}{rgb}{0, 0, 0.5}
\definecolor{darkblue}{rgb}{0, 0, 0.7}
\definecolor{purple}{rgb}{0.5, 0, 0.5}

\def\oneone{\rlap 1\mkern4mu{\rm l}}

\def\coeff#1#2{\relax{\textstyle \frac{#1}{#2}}\displaystyle}

\def\IR{\mathbb{R}}

\def\ZZ{\mathbb{Z}}

\def\cC{{\cal C}}

\def\cI{{\cal I}}

\def\cL{{\cal L}}

\def\cN{{\cal N}}
\def\cO{{\cal O}}

\def\nca{{\zeta}}
\def\ncb{{\xi}}

\def\nBPS#1{$\frac{1}{#1}$-BPS}

\numberwithin{equation}{section}

\begin{document}

\phantom{AAA}
\vspace{-10mm}

\vspace{1.9cm}

\begin{center}

{\huge {\bf   Localizing Momentum Waves on Mazes  }}\\

{\huge {\bf \vspace*{.25cm}  }}

\vspace{1cm}

{\large{\bf { Tobi Ramella $^{1}$  and  Nicholas P. Warner$^{1,2,3}$}}}

\vspace{1cm}

\centerline{$^1$Department of Physics and Astronomy}
\centerline{and $^2$Department of Mathematics,}
\centerline{University of Southern California,} 
\centerline{Los Angeles, CA 90089, USA}

\vspace{1cm}

$^3$Institut de Physique Th\'eorique, \\
Universit\'e Paris Saclay, CEA, CNRS,\\
Orme des Merisiers, Gif sur Yvette, 91191 CEDEX, France \\[12pt]

\vspace{10mm} 
{\footnotesize\upshape\ttfamily  ramella  @ usc.edu\,, \quad  warner @ usc.edu} \\

\vspace{1.5cm}
 
\textsc{Abstract}

\end{center}

\noindent We examine the  proposal  and analysis of \cite{Bena:2024qed} for putting momentum waves on mazes of intersecting M2 and M5 branes by computing a family of simple examples of momentum waves on pure M5 branes.  We re-analyze the BPS equations for the near-brane M5-P system with the prescribed supersymmetry structure of the full M2-M5-P system and show that the general linear system obtained in  \cite{Bena:2024qed} captures all of the null momentum waves on the simpler M5-P system.  Despite the absence of explicit M2 branes in our examples, we find that  the momentum waves still localize on an AdS$_3$  factor of the AdS$_7$ associated with M5 branes, showing that the supersymmetry prescription still reflects the M2-M5-P structure.  This momentum localization takes two forms:  explicit singular sources and smooth, localized bump functions reflecting  ``momentum migration'' that is a feature of other microstate  geometries. The simple, physical form of the momentum localization in our examples supports the broader proposal in  \cite{Bena:2024qed} that independent momentum waves can localize at each brane intersection.  We suggest further generalizations of momentum waves at brane intersections that could have important implications in holography.

\begin{adjustwidth}{3mm}{3mm} 
 
\vspace{-1.2mm}
\noindent

\end{adjustwidth}

\thispagestyle{empty}
\newpage


\tableofcontents

\section{Introduction}
\label{sec:Intro}
 
 Finding supergravity geometries that are dual to coherent families of black-hole microstates has been an enterprise that has spanned more than two decades (for recent reviews, see \cite{Bena:2022ldq,Bena:2022rna,Bena:2025pcy}).  The majority of such known  ``microstate geometries'' are supersymmetric, and there is a  huge variety of such solitonic solutions.  In retrospect, this is perhaps not so surprising in that a black hole has an outrageous number of microstates (compared a normal star, or even all the normal stars in a galaxy) and so one might expect a similarly outrageous number of semi-classical coherent states in the phase space of a black hole.  
 
Many examples of microstate geometries were found in five-dimensional \cite{Giusto:2004kj,Bena:2005va,Berglund:2005vb,Bena:2006is,Bena:2006kb,Bena:2007kg,Bena:2007qc,Bena:2008wt,Bena:2010gg,Gibbons:2013tqa,Bena:2011dd, Bena:2013dka, Bianchi:2016bgx, Bianchi:2017bxl, Heidmann:2017cxt, Bena:2017fvm, Avila:2017pwi,Tyukov:2018ypq,Warner:2019jll} and six-dimensional \cite{Shigemori:2013lta,Giusto:2013bda,Bena:2015bea,Bena:2016ypk,Bena:2017geu,Bena:2017upb,Bena:2017xbt,Ceplak:2018pws,Heidmann:2019xrd,Heidmann:2019zws,Walker:2019ntz,Ganchev:2021iwy}  supergravity, and the latter solutions have been mapped onto states of the D1-D5 CFT.  However, the number of such states, while extremely large, falls parametrically short of the black-hole entropy \cite{Shigemori:2019orj,Mayerson:2020acj}.    It became evident that to capture the majority of the microstructure one must find a way to access the twisted sectors of the dual CFT using fractionated branes in supergravity.   That is, one starts with two species of intersecting branes that form  a \nBPS{4}  system, or substrate, and then creates the  \nBPS{8} black-hole microstructure by adding a momentum charge.   If the original branes fractionate into one another then there are  $\sim N_1 N_2$ moduli describing all the intersections (where  $N_1$ and $N_2$ are the numbers of original branes).   The black-hole entropy, $S \sim \sqrt{N_1 N_2 N_P}$, then emerges from partitioning $N_P$ units of momentum into fractionated excitations.  This sort mechanism was the key to reproducing the black-hole entropy in  the first successful perturbative descriptions of black-hole entropy \cite{Strominger:1996sh,Maldacena:1997de,Dijkgraaf:1996cv}.    The idea proposed and developed in  \cite{Bena:2022wpl,Bena:2022fzf,Bena:2023rzm,Bena:2024qed,Bena:2024dre,Bena:2025hxt} was to see whether there were such constructions that could be given coherent descriptions within supergravity. 

The starting point of   \cite{Bena:2023rzm,Bena:2024qed,Bena:2024dre,Bena:2025hxt} is a substrate of intersecting M2 and M5 branes with the time, $t$, and one spatial direction, $y$, in common.  One can add a further set of M5 branes, usually denoted M5', that share $(t,y)$ but are otherwise transverse to the original M2 and M5 branes.  This does not break the supersymmetry any further: it is still a \nBPS{4} substrate.  The simplest corresponding geometries have $SO(4) \times SO(4)$ symmetry and have been extensively studied in \cite{Lunin:2007ab,Lunin:2007mj,Lunin:2008tf,Bena:2023rzm,Bena:2024dre,Bena:2025hxt}, and, as shown in \cite{Bena:2023rzm,Bena:2024dre,Bena:2025hxt}, have ``near-brane'' limits with   AdS$_3 \times$S$^3 \times$S$^3 \times \Sigma$ backgrounds,  constructed in \cite{Bachas:2013vza}.   

There is also an intriguing CFT that is the holographic dual to these geometries,  which have has been investigated from the perspective of the compactification to (as well as dualization within)  type II supergravity \cite{Boonstra:1998yu,Elitzur:1998mm,deBoer:1999gea,Gukov:2004ym,Tong:2014yna,Eberhardt:2017fsi,Eberhardt:2017pty,Eberhardt:2019niq,Witten:2024yod}.   While this CFT is not well understood, it is the theory that underlies the microstates of the \nBPS{8} M2-M5-P black hole.  One of the goals of the work in \cite{Bena:2022wpl,Bena:2022fzf,Bena:2023rzm,Bena:2024qed,Bena:2024dre,Bena:2025hxt} is to find supergravity descriptions of coherent excitations of this CFT in much the same way as superstrata provide coherent states of the D1-D5 CFT.  The advantage of studying the M2-M5-P system is that one can focus of individual brane intersections, attempt to map out the semi-classical picture for each such intersection and then, hopefully, assemble a more complete picture of momentum waves on fractionated branes.

A critical first step in this program was made in \cite{Bena:2024qed}, where it was shown how to start from a \nBPS{4}  M2-M5-M5' substrate and add  fluxes that carry momentum charge thereby creating new \nBPS{8} M2-M5-M5'-P solutions.  The challenge of the  \nBPS{4}  M2-M5-M5' substrate is that the solution is governed by a  non-linear equation that is akin to a complicated generalization of the Monge-Amp\`ere equation.  In \cite{Lunin:2007mj} (see Sections 4.5 and 5.1) it was argued, using perturbation theory, that once one has specified a  brane distribution through boundary conditions and sources, there is a unique solution to this equation.  Moreover, in the near-brane limit studied in \cite{Bachas:2013vza,Bena:2023rzm,Bena:2024dre,Bena:2025hxt}, the system of equations governing the solution becomes linear and is far more solvable.  

One of the surprising results of  \cite{Bena:2024qed} was that, even if one starts from a generic M2-M5-M5' substrate governed by its non-linear equation, the additional fluxes and momentum waves are determined by a {\it linear} system of equations in the fixed background of the \nBPS{4}  M2-M5-M5' substrate. This bodes very well for the holography of such waves because it means that the phase space of these momentum excitations does not involve complicated, non-linearly constrained excitations: Once one has the  \nBPS{4} substrate, one can apply the superposition principle to all the momentum waves.  This was also a crucial feature of all the momentum excitations of superstrata, and it led to a complete  holographic dictionary for the superstratum states in the CFT.
 
The remarkable aspect of the results of \cite{Bena:2024qed} was that the momentum waves, and fluxes that carry them, are governed by a linear system, however these results were obtained for a generic M2-M5-M5' substrate and the properties of these solutions remained unclear.   The purpose of this paper is to examine these momentum waves in a very simple example and explore some important physical questions about the waves and the fluxes that carry them.  First and foremost, we show that the linear system of equations for these excitations do indeed lead to some very simple families of physically sensible solutions.   We also explore the extent to which the waves and fluxes localize around the branes.  The larger picture of black-hole microstructure  requires that  there should be an independent set of momentum waves at each brane intersection, and for this to be realized, the momentum waves should be localizable around individual branes.  Moreover, for the momentum waves to represent  states of the holographic CFT, the modes must fall off sufficiently fast in the AdS throats created by the branes.  
 
We focus on pure M5 branes because these are the heaviest of the branes, and locally dominate the substrate solutions analyzed in \cite{Bena:2023rzm,Bena:2024qed,Bena:2024dre,Bena:2025hxt}, and we work in the near-brane limit and so the substrate is simply Poincar\'e  AdS$_7 \times S^4$.  However,  we start with a background  adapted to the presence of M2 branes: the Poincar\'e  slices are  decomposed into an $\IR^{1,1}$ defined by the common M2 and M5 directions, $(t,y)$, and an $\IR^4$, with Euclidean coordinates $\vec u$,  defining the directions inside the M5's but transverse to the M2's.  The $\IR^5$ transverse to the M5's  decomposes into $\IR \times \IR^4$, with coordinates  $(z, \vec v)$, where $\IR$ is the remaining M2 direction and the $\IR^4$ are the directions transverse to both the M2's and M5's.  We repeat the analysis of \cite{Bena:2024qed} for the pure M5 background, but we still impose the supersymmetries of the M2-M5-P system, and solve, {\it ab initio}, the  BPS equations and the equations of motion.   Given the absence of M2 branes, one might, perhaps, expect a broader class of momentum wave solutions, however we find that the solutions for the M5-brane alone are entirely captured by the general linear system obtained  for the M2-M5-P system in  \cite{Bena:2024qed}.

The work on momentum waves discussed in \cite{Bena:2024qed} assumed an $SO(4) \times SO(4)$ symmetry, which means that the solutions were constrained to be functions of $u = |\vec u|$, $v = |\vec v|$  and $z$.  We strongly suspect that this high level of symmetry is not required and that there will be waves that oscillate with non-trivial harmonics on the $S^3 \times S^3$ in $\IR^4 \times \IR^4$.  Here, because the coordinates $(z, \vec v)$ combine to describe an $\IR^5$, we can use this dependence to explore modes on the $S^4$ that surround the M5's. The AdS$_7$ radius is then defined by $r = \sqrt{v^2 +z^2}$, and the locus of the M5 branes is at the bottom of the  AdS$_7$ throat, $r=0$.   The locus of any would-be M2 branes is defined by $u=0$ and $v=0$. 

The geometry decomposes similarly: the self-similar structure of the M2-M5 spikes \cite{Bena:2023rzm,Bena:2025hxt} defines an AdS$_3$ corresponding to the $u=0$ locus within the M5 branes.  For excitations of this system, there is a very important distinction between  AdS$_7$ and  AdS$_3$.   Ultimately, for smooth microstate geometries, we may want to consider the global forms of these AdS geometries, and these have spatial sections of $S^5$ and $S^1$.   The $S^5$ has gapped excitations because of its non-trivial curvature, whereas $S^1$ is flat and can be thought of as  a periodic identification of $\IR$ in Poincar\'e AdS.  We therefore expect the waves that localize at $u=0$ to describe states that are qualitatively different from those described by waves that explore the larger M5 world volume.  We are therefore going to be particularly interested in how the waves and fluxes localize both as a function of $r$ and as a function of $u$. 

The obvious form of localization are the waves and fluxes that have singular sources at $r=0$, or at $u=0$, or both.  However, it was noted in \cite{Bena:2025uyg} that micorstate geometries often exhibit ``momentum migration.''  That is, momentum waves, and the fluxes that carry them, can  move off the brane sources while being constrained to lie close to them.  In such solutions, the gravitational back-reaction actually causes the  momentum modes to vanish at the loci of the original branes  and, instead, these modes  become localized, by smooth bump functions, in regions close to the original brane locus.  Here we will find both forms of localization in the momentum excitations.

 In Section \ref{sec:3dSugr}  we review the construction in  \cite{Bena:2024qed}  of momentum-carrying M2-M5-P system and the layered BPS structure governing the momentum sector.   In Section \ref{sec:M5back} we specialize to the near-brane M5 background and repeat the analysis of BPS momentum waves given in   \cite{Bena:2024qed}.   We find that even in this simplified background, the BPS momentum waves are captured by the linear system of equations obtained  in \cite{Bena:2024qed}, and we are reduced to solving Laplace and Poisson equations on  AdS$_7 \times S^4$.  In these computations, we find that the self-similar scaling structure that leads to the AdS$_3$ sections of M2-M5 spikes naturally emerges as a very useful organizational principle and we catalog the general properties of various scaling momentum-wave solutions.   In Section \ref{sec:Examples} we construct a range of examples revealing the different types of momentum localization, including some smooth ``bump-function'' solutions.  Section~\ref{sec:Conclusions} contains our final remarks and conclusions.

\section{Momentum on generic M2-M5-M5's intersections}
\label{sec:3dSugr}

\subsection{The \nBPS{8} solution}
\label{ss:4susysolution}

The metric is given by:
\begin{equation}
\begin{aligned}
ds_{11}^2 ~=~  e^{2  A_0}\, \bigg[ &  d \ncb \, \bigg( P \,  d  \ncb    ~+~ 2\, \bigg( \frac{ d \nca}{F(\ncb) } ~+~ k \,  \sigma_3  \bigg)\bigg)~+~ e^{-3  A_0} \, (-\partial_z w )^{-\frac{1}{2}}\, ds_4^2 \\ & ~+~ e^{-3  A_0} \, (-\partial_z w )^{\frac{1}{2}}\,  {ds'}_4^2
 ~+~  (-\partial_z w ) \, \big( dz ~+~(\partial_z w )^{-1}\,  (\partial_u w ) \, d u \big)^2  \bigg]\,,
\end{aligned}
 \label{11metric}
\end{equation}
where  ${ds}_4^2$ and ${ds'}_4^2$ are flat Euclidean metrics in $\IR^4$ written in terms of left-invariant $1$-forms: 
\begin{equation}
 ds_{4}^2 ~=~     du^2  ~+~ \coeff{1}{4} \, u^2 \, (\sigma_1^2 +\sigma_2^2+ \sigma_3^2 \big)  \,, \qquad {ds'}_4^2  ~=~  dv^2  ~+~ \coeff{1}{4} \, v^2 \, \,({\sigma'}_1^2 +{\sigma'}_2^2+{\sigma'}_3^2\big) \,.
 \label{4metrics}
\end{equation}
Indeed, we take 
\begin{equation}
\begin{aligned}
\sigma_1 ~=~  &   \cos \varphi_3 \, d \varphi_1 ~+~ \sin \varphi_3 \sin \varphi_1\, d \varphi_2  \,, \\ 
\sigma_2 ~=~  &   \sin \varphi_3 \, d \varphi_1 ~-~ \cos \varphi_3 \sin \varphi_1\, d \varphi_2  \,, \\ 
\sigma_3 ~=~  &    d \varphi_3 ~+~ \cos  \varphi_1\, d \varphi_2  \,,
\end{aligned}
 \label{1forms}
\end{equation}
with the similar expressions for the ${\sigma'}_i$, but with $\varphi_j \to {\varphi'}_j$.   This corresponds to  Choice (ii)  in \cite{Bena:2024qed}.  The coordinates $\nca$ and $\ncb$ are null coordinates related to the common Poincar\'e time and space coordinates $(t,y)$, along the branes by:
\begin{equation}
\nca ~=~ \coeff{1}{\sqrt{2}} \, (y+t) \,, \qquad \ncb ~=~ \coeff{1}{\sqrt{2}} \, (y-t) \,.
 \label{tynull}
\end{equation}

There is an inherent danger in making Choice (ii)  of \cite{Bena:2024qed}:  the polarization vector, $ k \sigma_3$, will generically become singular in the limit $u \to 0$ where the $S^3$ in the first $\IR^4$ collapses to a point. We will encounter this later in our analysis.  Choice (i)  of \cite{Bena:2024qed} avoids this problem by taking the polarization vector to point along a flat, uncompactified direction in the $\IR^4$.  However, we wish to make contact with the results of \cite{Bachas:2013vza,Bena:2025hxt} and this requires that we preserve the $S^3$ in the first $\IR^4$ factor, and hence  make Choice (ii). 

The metric Ansatz involves five, as yet, arbitrary functions.  Four of these,  $P, k, w$ and $A_0$, are functions of $(u,v,z)$ and are determined by the BPS equations and the equations of motion.  The remaining function, $F(\ncb)$, is unconstrained and determines the wave profile.   

To analyze the supersymmetries, one introduces a set of frames 
\begin{equation}
\begin{aligned}
e^0 ~=~  & \frac{e^{A_0} }{\sqrt{P} }\,\bigg( \frac{ d \nca}{F(\ncb) } + k \,  \sigma_3  \bigg)  \,, \qquad e^1~=~    \frac{e^{A_0} }{\sqrt{P} }\, \bigg(P \,  d  \ncb    +   \frac{ d \nca}{F(\ncb) } + k \,  \sigma_3 \bigg)   \,, \\ e^2 ~=~ &  e^{A_0} (-\partial_z w )^{\frac{1}{2}} \, \Big( dz ~+~(\partial_z w )^{-1}\,  \big (\vec \nabla_{\vec u} \, w \big)  \cdot  d \vec u \Big) \,, \\
\qquad e^3 ~=~& e^{- \frac{1}{2} A_0} \, (-\partial_z w )^{-\frac{1}{4}}\,   du  \,,   \qquad e^4 ~=~   e^{- \frac{1}{2} A_0} \, (-\partial_z w )^{\frac{1}{4}}\,  dv \,,\\
e^{i+4}~=~  &   \coeff{1}{2} \, e^{- \frac{1}{2} A_0} \, (-\partial_z w )^{-\frac{1}{4}}\, u \, \sigma_i  \,, \qquad e^{i+7} ~=~   \coeff{1}{2} \,  e^{- \frac{1}{2} A_0} \, (-\partial_z w )^{\frac{1}{4}}\, v \, {\sigma'}_i \,,   \qquad {i = 1,2,3}       \,.
\end{aligned}
 \label{11frames}
\end{equation}

The \nBPS{8} solution carries M2, M5 and momentum (P) charge, and its supersymmetries are required to satisfy the projection conditions (in the frame basis (\ref{11frames})): 
\begin{equation}
 \Gamma^{01} \, \varepsilon  ~=~ - \varepsilon \,,   \qquad   \Gamma^{012} \, \varepsilon  ~=~ - \varepsilon \,,   \qquad  \Gamma^{013567} \, \varepsilon  ~=~\varepsilon \,.
 \label{projs3}
\end{equation}
Recalling that, in eleven dimensions, one has $ \Gamma^{0123456789\,10} =  \oneone$, one sees that (\ref{projs3})  implies
\begin{equation}
 \Gamma^{01489\,10} \, \varepsilon  ~=~ \varepsilon \,,
 \label{projs2}
\end{equation}
and hence one can add another set of M5 branes along the directions $01489\,10$ without breaking supersymmetry any further.  As in \cite{Bena:2024qed}, we will denote this second possible set of branes by M5', but we will not actively include sources for such branes.

In the \nBPS{8} solution, the Maxwell potential takes the form:
\begin{equation}
\begin{aligned}
 C^{(3)} ~=~ &  -  e^0 \wedge e^1 \wedge e^2 \\ & - \frac{1}{8} \,\bigg(\frac{u^3\, \partial_u w}{\partial_z w}\bigg)\,  \sin \varphi_1 \, d\varphi_1 \wedge d\varphi_2  \wedge d\varphi_3
- \frac{1}{8} \, (v^3 \,\partial_v w)\, \sin {\varphi'}_1 \, d{\varphi'}_1 \wedge d{\varphi'}_2  \wedge d{\varphi'}_3 \\
 & +  \frac{1}  {u^4 \sqrt{P}\, (-\partial_z w)^{\frac{1}{2}}} \, \big(\partial_z p \big)  \, (e^1-e^0)\wedge \big( e^3\wedge e^7~-~ e^5\wedge e^6 \big)   \,, 
 \end{aligned}
 \label{C3res1}
\end{equation}
where $p$ is another function of $(u,v,z)$ that will ultimately be determined by a linear, homogeneous, second-order differential equation.  

Note that the last term in (\ref{C3res1}) may be written as 
\begin{equation}
 C^{(3)}_{wave} ~=~   \frac{\big( \partial_z p \big) }  {4\, u^4  (-\partial_z w)} \, d\xi \wedge \big( 2\, u \,du \wedge \sigma_3~-~u^2 \, \sigma_1 \wedge\sigma_2 \big)   \,, 
 \label{C3part1}
\end{equation}
which does not involve the function, $P$.  As we will see, $k$ is also determined entirely by $p$, so that $P$'s only role is as the coefficient of $d\ncb^2$.  In particular, despite the appearance of $\sqrt{P}$ in the  frames, the metric and flux do not impose any constraints on the sign of $P$.  Indeed the metric is Lorentzian for any sign of $P$.

\subsection{The \nBPS{4} substrate}
\label{ss:substrate}

To describe the BPS system and equations of motion, it is convenient to start from the  \nBPS{4} substrate solution in which the momentum is turned off. That is, one takes $P \equiv k \equiv p \equiv 0$ and $F(\ncb) \equiv 1$, so that the metric and flux reduce to the solution described in \cite{Lunin:2008tf,Bena:2023rzm,Bena:2024dre,Bena:2025hxt} with
\begin{equation}
\begin{aligned}
ds_{11}^2 ~=~  e^{2  A_0}\, \Big[ -dt^2 &~+~ dy^2 ~+~ e^{-3  A_0} \, (-\partial_z w )^{-\frac{1}{2}}\, ds_4^2  ~+~ e^{-3  A_0} \, (-\partial_z w )^{\frac{1}{2}}\,  {ds'}_4^2 \,    \\
  &  ~+~  (-\partial_z w ) \, \big( dz ~+~(\partial_z w )^{-1}\,   (\partial_u \, w )   \,d  u \big)^2  \Big]\,,
\end{aligned}
 \label{11metric-subs}
\end{equation}
\begin{equation}
 C^{(3)} ~=~  -  e^0 \wedge e^1 \wedge e^2 ~+~\bigg(\frac{u^3 \,\partial_u w}{\partial_z w}\bigg)\,  \sigma_1 \wedge \sigma_2  \wedge \sigma_3
 + (v^3 \,\partial_v w)\,{\sigma'}_1 \wedge {\sigma'}_2  \wedge {\sigma'}_3  \,. \label{C3res2}
\end{equation}

As was discussed in \cite{Lunin:2008tf,Bena:2023rzm}, the solution can be reduced to solving a Monge-Amp\`ere-like equation for a pre-potential. $G_0$:
\begin{equation}
 {\cal L}_{v} G_0 ~=~  ({\cal L}_{ u} G_0)\,(\partial_z^2 G_0) ~-~ ( \partial_u \partial_z G_0)^2\,, 
 \label{maze-eq}
\end{equation}
where  ${\cal L}_{u}$ and ${\cal L}_{v}$ are the radial Laplacians on the respective $\IR^4$'s:
\begin{equation}
 {\cal L}_{u} H  ~\equiv~   \frac{1}{u^3}\, \partial_u \big( u^3  \partial_u H \big)  \,,  \qquad  {\cal L}_{v} H  ~\equiv~   \frac{1}{v^3}\, \partial_v \big(v^3 \partial_v H \big)  \,.
 \label{Laps}
\end{equation}
The metric functions can then be recovered from:
\begin{equation}
w ~\equiv~ \partial_z G_0  \,, \qquad   e^{-3  A_0} \, (-\partial_z w )^{\frac{1}{2}} ~=~ {\cal L}_{v}  G_0 \,.
 \label{solfns}
\end{equation}

The important point is that adding a momentum wave does not change these functions, or their equations: $A_0$ and $w$ remain the same in the full \nBPS{8}  solution.

\subsection{The next layers of the BPS system}
\label{ss:BPSlayers}

To define the full BPS system it is convenient to introduce the Laplacian, $\hat {\cal  L}$, for the  substrate metric (\ref{11metric-subs}) acting on a  function  $H$ that only depends on $(u,v,z)$.   Using the equations for $w$ and $A_0$,  one can simplify the Laplacian  to obtain the following operator: 
\begin{equation}
\begin{aligned}
{\cal L} ( H) ~\equiv~ &e^{- A_0}\,(-\partial_z w )^{-\frac{1}{2}} \,   \hat {\cal L} ( H)  \\
 ~=~   &   \bigg[  \frac{1}{u^3}\, \partial_u \big( u^3  \partial_u H \big) ~+~  \frac{1}{ (- \partial_z w)}\,  \frac{1}{v^3} \partial_v \big( v^3  \partial_v H \big) ~+~2\, \frac{( \partial_u w)}{ (- \partial_z w)} \, \partial_u \partial_z H     \\
  & ~+~  \Big((-\partial_z  w)^{-\frac{3}{2}} \,e^{-3  A_0}~+~  (-\partial_z  w)^{-2} \,  (\partial_u w)^2\big)\Big)\, \partial_z^2 H \bigg] \,,
\end{aligned}
 \label{Lap1}
\end{equation}
It is interesting to note that one could also have replaced $\hat {\cal  L}$ by the Laplacian for the final metric (\ref{11metric}) because {\it on functions of $(u,v,z)$ alone}, these two Laplacians agree. Here, however, we wish to emphasize that $\hat {\cal L}$, and hence ${\cal L}$, is a linear operator on a known substrate background that does not depend upon the functions we are trying to determine. 

\subsubsection{The flux function}
\label{ss:fluxfn}

It was shown in \cite{Bena:2024qed} that the flux-function, $p$, satisfies:
\begin{equation}
{\cal L} \bigg( \frac{p}{u^4}\bigg) ~-~ \frac{8 }{u^2}\, \frac{p}{u^4}~=~  0 \,,
 \label{peqn}
\end{equation}
which, as promised, is a linear, homogeneous, second-order differential equation coming from the Laplacian on the substrate metric.   The polarization function, $k$, is then obtained from $p$ via \cite{Bena:2024qed}\footnote{There is a typographical error in  \cite{Bena:2024qed} in which $\partial_v  p$ should have been  $\partial_z  p$ and this propagated into  subsequent equations.  We have corrected that error here.}:
\begin{equation}
k ~=~\frac{1}{2\, u^3} \, \bigg(\partial_u p~-~ \frac{(\partial_u w)}{(\partial_z w)}\, \partial_z  p \bigg)    \,.
 \label{kform1}
\end{equation}

The {\it new} non-zero  field strengths,  in frame indices, are:
\begin{equation}
\begin{aligned}
F_{0237} =&  -F_{1237} ~=~  b_1 \,, \qquad   F_{0347}=  -F_{1347} ~=~  b_2 \,, \qquad  F_{0247} =  -F_{1247} ~=~ \frac{1}{2}\,  \frac{e^{2 A_0}}{\sqrt{P}}\,   (\partial_v k) \,,   \\ F_{0256} = & -F_{1256} ~=~  -b_1~+~    \frac{e^{2 A_0}}{u\, \sqrt{P}}\,   \bigg[ \bigg( (-\partial_z w )^{\frac{1}{2}}\,  \partial_u k+ \frac{\partial_u w}{(-\partial_z w )^{\frac{1}{2}}}\,  \partial_z k\bigg)  ~+~  \frac{2}{u}\,(-\partial_z w )^{\frac{1}{2}}\,k \,  \bigg] \,,
\end{aligned}
 \label{fluxfunctions2}
\end{equation}
where $b_1, b_2$ are given by:
\begin{equation}
\begin{aligned}
 b_1 ~=~ & \frac{ e^{ {2A_0}}}{u^4\,  \sqrt{P}} \,   \bigg[\,2\,  u^3 \,   \bigg( (-\partial_z w )^{\frac{1}{2}}\,  \partial_u k+ \frac{\partial_u w}{(-\partial_z w )^{\frac{1}{2}}}\,  \partial_z k\bigg) ~-~ e^{-3 A_0}\,  \,\partial_z \bigg( \frac{\partial_z p}{\partial_z w}\bigg) \,\bigg]   \,, \\
 b_2  ~=~ &  \frac{ e^{ {\frac{1}{2} A_0}}}{u^4\,  \sqrt{P}} \, (-\partial_z w)^{\frac{1}{4}} \,\partial_v \bigg( \frac{\partial_z p}{\partial_z w}\bigg)\,.
 \end{aligned}
 \label{bres1}
\end{equation}
%

\subsubsection{The momentum density}
\label{ss:momdens}

 The momentum density function, $P(u,v,z)$, is determined by the magnetic flux  sources.
\begin{equation}
{\cal L} \big(P\big)  ~=~  s  \,,
 \label{Peqn}
\end{equation}
where:
\begin{equation}
\begin{aligned}
s  ~\equiv~ - 8\,e^{- A_0}\,(-\partial_z w )^{-\frac{1}{2}} \,\bigg[&  \big(\sqrt{P} \, b_2\big)^2 
~+~  \bigg(\big(\sqrt{P} \, b_1\big) ~-~ \frac{2\, e^{2 A_0}}{u^2}\,(-\partial_z w )^{\frac{1}{2}}\,k\, \bigg)  \\ &  \times \bigg(\big(\sqrt{P} \, b_1\big) ~-~ \frac{e^{2 A_0}}{u} \,\Big((-\partial_z w )^{\frac{1}{2}}\,\partial_u k~+~ (-\partial_z w )^{-\frac{1}{2}}\,(\partial_u w)\,\partial_z k\,\Big) \bigg)\bigg]\,. 
\end{aligned}
 \label{source1}
\end{equation}
%

\section{The momentum waves on an M5-brane background}
\label{sec:M5back}

\subsection{The M5 substrate}
\label{ss:substrateM5}

We now specialize to the near-brane limit of a pure M5-brane background, and the easiest way to do this is to start from the corresponding solution to (\ref{maze-eq}):
\begin{equation}
G_0 ~=~ - \bigg( \alpha^3\, \frac{\sqrt{v^2 + z^2}}{v^2} ~+~ \frac{1}{8} \, u^2 \bigg)\,, 
 \label{maze-sol1}
\end{equation}
where $\alpha >0$ is a constant.
This leads to:
\begin{equation}
w ~=~ -  \frac{\alpha^3\,  z }{v^2 \sqrt{v^2 + z^2}} \,, \qquad \partial_z w ~=~ -  \frac{\alpha^3 }{(v^2 + z^2)^{3/2}} \,, \qquad e^{2 A_0} ~=~\frac{\sqrt{v^2 + z^2}} {\alpha } \,. 
 \label{maze-sol2}
\end{equation}
The metric on the substrate (\ref{11metric-subs})  is then
\begin{equation}
\begin{aligned}
ds_{11}^2 ~=~  \alpha^2\, \bigg[\frac{\sqrt{v^2 + z^2}} {\alpha^3 }\, \bigg( -dt^2 &~+~ dy^2 ~+~  du^2  ~+~ \coeff{1}{4} \, u^2 \, (\sigma_1^2 +\sigma_2^2+ \sigma_3^2 \big) \bigg) \\
& ~+~ \frac{1}{(v^2 + z^2)}\big(   dz^2  ~+~ dv^2  ~+~ \coeff{1}{4} \, v^2 \, \,({\sigma'}_1^2 +{\sigma'}_2^2+{\sigma'}_3^2\big) \big)   \bigg]\,.
\end{aligned}
 \label{11metric-subs-simp1}
\end{equation}
Setting $z = r \cos \theta, v = r \sin \theta$ leads to:
\begin{equation}
ds_{11}^2 ~=~\frac{r} {\alpha}\, \Big( -dt^2 ~+~ dy^2 ~+~  du^2  ~+~ \coeff{1}{4} \, u^2 \, (\sigma_1^2 +\sigma_2^2+ \sigma_3^2 \big) \Big) \\
~+~ \frac{\alpha^2}{r^2 }\big(   dr^2  ~+~ r^2 \, d\Omega_4^2 \big) \,,
 \label{11metric-subs-simp2}
\end{equation}
where 
\begin{equation}
d\Omega_4^2 ~\equiv~ d\theta^2  ~+~ \coeff{1}{4} \, \sin^2 \theta \, \,({\sigma'}_1^2 +{\sigma'}_2^2+{\sigma'}_3^2\big)
 \label{4sphere-met}
\end{equation}
is the metric on a unit $S^4$.

This corresponds to the standard near-brane limit of a stack of M5 branes:
\begin{equation}
ds_{11}^2   ~=~H( r)^{-\frac{1}{3}}  \, \eta_{\mu \nu}\, d x^\mu  dx^\nu ~+~ H( r)^{\frac{2}{3}}  \, \big( \, d r^2 ~+~ r^2  \, d \Omega_4^2 \, \big)     \,, \qquad H( r) = \frac{Q_5 }{ r^3} \,,
 \label{M5stack1}
\end{equation}
and so the brane charge is:
\begin{equation}
Q_5  ~=~ \alpha^3 \,.
 \label{Q5}
\end{equation}

The gauge fields are given by:
\begin{equation}
\begin{aligned}
 C^{(3)} ~=~&  -  e^0 \wedge e^1 \wedge e^2 ~+~\bigg(\frac{u^3 \,\partial_u w}{\partial_z w}\bigg)\,  \sigma_1 \wedge \sigma_2  \wedge \sigma_3
 + (v^3 \,\partial_v w)\, {\sigma'}_1 \wedge {\sigma'}_2  \wedge {\sigma'}_3  \\
 ~=~&  -  dt \wedge dy \wedge dz  ~-~ \frac{\alpha^3}{8} \, \cos \theta \, (3 -\cos^2 \theta)\, \sin {\varphi'}_1 \, d{\varphi'}_1 \wedge d{\varphi'}_2  \wedge d{\varphi'}_3   \,. \label{C3simp1}
\end{aligned}
\end{equation}
so that 
\begin{equation}
\begin{aligned}
 d C^{(3)} ~=~ \frac{\alpha^3}{8} \, \sin^3 \theta \, \sin {\varphi'}_1 \, d{\theta}\wedge d{\varphi'}_1 \wedge d{\varphi'}_2  \wedge d{\varphi'}_3   \,,
 \label{F4simp1}
\end{aligned}
\end{equation}
which is the volume form on $S^4$, and therefore consistent with  (\ref{Q5}).

Since $\partial_u w =0$, the form of the polarization function, $k$, (\ref{kform1}) also becomes much simpler: 
\begin{equation}
k ~=~\frac{1}{2\, u^3} \,  \partial_u p   \,.
 \label{kform2}
\end{equation}

One should also note that the metric (\ref{11metric-subs-simp2}) can be put in the standard form of Poincar\'e AdS$_7 \times S^4$ by setting $r =2 \alpha^{3/2} \mu^2$:
\begin{equation}
ds_{11}^2 ~=~ \alpha^2\, \bigg[ 4\, \bigg( \frac{d \mu^2}{\mu^2} ~+~ \mu^2 \Big( -dt^2 +  dy^2 + du^2  + \coeff{1}{4} \, u^2 \, (\sigma_1^2 +\sigma_2^2+ \sigma_3^2 \big) \Big) \bigg)\\
~+~   d\Omega_4^2 \bigg] \,,
 \label{11metric-AdS}
\end{equation}
This leads to the expected scale invariance $\mu \to \lambda  \mu$, $(t, y, u) \to \lambda^{-1} (t, y, u)$.  In terms of the $(u,v,z)$ coordinates, this AdS scale invariance is 
\begin{equation}
  u  \to \lambda^{-1}\, u \,, \qquad   v  \to \lambda^2 \, v \,, \qquad  z  \to \lambda^2 \, z   \,.
 \label{scaling1}
\end{equation}
This scaling behavior will provide an invaluable scheme for organizing the momentum-wave solutions.

\subsection{The \nBPS{8} momentum wave}
\label{ss:M5wave}

Rather than use the result of \cite{Bena:2024qed} that was described in  Section \ref{ss:4susysolution}, we start, {\it ab initio} from the M5 brane background of Section \ref{ss:substrateM5}, and use the momentum wave Ansatz  (\ref{11frames}) specialized to the M5 substrate.  That is, we use the frames:
\begin{equation}
\begin{aligned}
e^0 ~=~  &\frac{1}{\sqrt{\alpha} } \, \frac{ (v^2 +z^2)^{\frac{1}{4}}}{\sqrt{P} }\,\bigg( \frac{ d \nca}{F(\ncb) } + k \,  \sigma_3  \bigg)  \,, \qquad e^1~=~    \frac{1}{\sqrt{\alpha} } \, \frac{ (v^2 +z^2)^{\frac{1}{4}}}{\sqrt{P} }\, \bigg(P \,  d  \ncb    +   \frac{ d \nca}{F(\ncb) } + k \,  \sigma_3 \bigg)   \,, \\ e^2 ~=~ &\frac{ \alpha \,  dz }{\sqrt{ v^2 +z^2} }  \,,  \qquad e^3 ~=~\frac{1}{\sqrt{\alpha} } \, (v^2 +z^2)^{\frac{1}{4}}\,   du  \,,   \qquad e^4 ~=~  \frac{ \alpha \,  dz }{\sqrt{ v^2 +z^2} }  \,  dv \,,\\
e^{i+4}~=~  & \frac{1}{2\,\sqrt{\alpha} } \,u\, (v^2 +z^2)^{\frac{1}{4}}\,    \sigma_i  \,, \qquad e^{i+7} ~=~ \frac{\alpha \,  v }{2 } \, \frac{ 1 }{\sqrt{ v^2 +z^2} }  \, {\sigma'}_i \,,   \qquad {i = 1,2,3}       \,.
\end{aligned}
 \label{11framesM5}
\end{equation}

We  then made a completely general Ansatz for the gauge fields consistent with the symmetries and, as in  \cite{Bena:2024qed}, and then we solved the BPS equations and the equations of motion.  The goal was  to see if there are any other solutions to this system given that we have a much simpler substrate.  However, we found that the only  solutions were those anticipated in   \cite{Bena:2024qed}.  That is, one must first solve (\ref{peqn}) and then determine the momentum density function, $P(u,v,z)$, from  (\ref{Peqn}).  

For the M5-brane substrate this becomes
\begin{equation}
{\cal L} \bigg( \frac{p}{u^4} \bigg) ~=~  {\cal L}_u  \bigg( \frac{p}{u^4} \bigg) ~+~\frac{(v^2 +z^2)^{\frac{3}{2}}}{\alpha^3} \,  \bigg[ \partial_z^2 \bigg( \frac{p}{u^4} \bigg) ~+~{\cal L}_v  \bigg( \frac{p}{u^4} \bigg) \bigg]~=~  \frac{8\,p}{u^6} \,,
 \label{peqn-simp1}
\end{equation}
where  ${\cal L}_u$ and ${\cal L}_u$ are given by (\ref{Laps}).   This can be rewritten as
\begin{equation}
{\cal L}_u  \bigg( \frac{p}{u^4} \bigg) ~+~\frac{r^3}{\alpha^3} \,  \bigg[\frac{1}{r^4} \, \partial_r  \bigg(r^ 4\, \partial_r \bigg( \frac{p}{u^4} \bigg) \bigg)  ~+~\frac{1}{r^2 \, \sin^3 \theta } \, \partial_\theta \bigg(\sin^3 \theta \, \partial_\theta \bigg( \frac{p}{u^4} \bigg) \bigg)  \bigg]~=~  \frac{8\,p}{u^6} \,.
 \label{peqn-simp3}
\end{equation}

It is important to note that the Laplacian, defined by (\ref{Lap1}),  is given by  (\ref{peqn-simp1})  in the M5 background,  and it scales under (\ref{scaling1}) according to 
\begin{equation}
{\cal L} ~\to~  \lambda^{2}\, \cal L  \,.
 \label{scaling2}
\end{equation}
As we will see, in simple solutions, the source, $s$,  (\ref{source1}), in  (\ref{Peqn}),  has a fixed scaling dimension and this is very useful in determining the form  of the function, $P$.

\subsection{Spherically symmetric, scaling solutions}
\label{ss:simpsol2}

Introduce the scale-invariant variable 
\begin{equation}
x  ~\equiv~   \frac{4\, \alpha^3}{u^2 \, \sqrt{v^2 +z^2} }\,, 
\label{xdefn}
\end{equation}
and make an Ansatz for solutions of (\ref{peqn-simp1}):
\begin{equation}
p(a;u, x)  ~=~ u^{a + 4} \,  F(a;x)   \,, 
\label{pAnsatz1}
\end{equation}
for some parameter, $a$.  One then finds that $F$ must satisfy the hypergeometric equation: 
\begin{equation}
x(1+x) \, \frac{d^2F}{dx^2}   ~-~ (a\,x +2) \, \frac{dF}{dx} ~+~ \bigg( \frac{a}{2}+2 \bigg) \bigg( \frac{a}{2} -1\bigg)  \, F~=~ 0 \,,
\label{Feqn}
\end{equation}
whose solution is:
\begin{equation}
F(a;x)     ~=~ x^3 \ {}_{2} F_1 \bigg(1- \frac{a}{2} , 4- \frac{a}{2}  , 4; -x \bigg)\,.
\label{Fsol1}
\end{equation}
One can use the properties of hypergeometric functions to obtain an alternative form for the solution: 
\begin{equation}
F(a;x)     ~=~(1+x)^{a-1}\,  x^3 \  {}_{2} F_1 \bigg(3+ \frac{a}{2} , \frac{a}{2}  , 4; -x \bigg)\,.
\label{Fsol2}
\end{equation}

For $a$ even and positive, the solutions (\ref{Fsol1}) are   Jacobi Polynomials (as functions of $1-2x$) of degree $\frac{a}{2} -1$ multiplied by  $x^3$.  The first few functions are
\begin{equation}
\begin{aligned}
F(2;x)  =  x^3  \,, \quad  F(4;x)    = \coeff{1}{2} \,  x^3 \,(x+2) \,, \quad F(6;x)    =\coeff{1}{10} \,  x^3 \, (x^2+5 x +10)  \,,   \dots 
\end{aligned}
\label{psol1}
\end{equation}
The function $p(a;u, x)$, is  finite as $u,v \to \infty$ and as $ u \to 0$, but becomes progressively more singular as $r \to 0$.

Henceforth we will take $a$ to be even because the series expansions of the hypergeometric terms contain $\Gamma$-functions that cause  many terms to vanish at particular values of even $a$,  This makes the asymptotic behavior much richer.

For $a \le 0$ and even one has, from (\ref{Fsol2}):
\begin{equation}
F(a;x)    ~=~\frac{x^3 }{(1+x)^{1-a}} \, Q(x)\,,
\label{Fsol3}
\end{equation}
where $Q(x)$ is, once again, a Jacobi Polynomial with argument $1- 2x$. However the degree of the polynomial jumps at $a= -6$.  The first few such functions are:
\begin{equation}
\begin{aligned}
F(0;x)  & =\frac{x^3 }{(1+x)} \,, \quad F(-2;x)    =\frac{x^3\,(2+x) }{2\,(1+x)^3} \,, \quad F(-4;x)    =\frac{x^3\,(x^2 +5x +10) }{10\,(1+x)^5} \,, \\ 
 F(-6;x)   & =\frac{x^3}{(1+x)^7}  \,, \quad F(-8;x)   =\frac{x^3\,(1-x) }{(1+x)^9}  \,, \quad F(-10;x)=\frac{x^3\,(2x^2 - 5x +2) }{2 \, (1+x)^{11}} \,, \dots 
\end{aligned}
\label{psol2}
\end{equation}
where, for $a \le -6$, the Jacobi polynomial factor in the numerator has degree $-(a+6)/2$.  These functions are all vanish as $u,v \to \infty$ and are finite as $ u \to 0$. As $r \to 0$ (with $u$ finite) one has:
\begin{equation}
\begin{aligned}
p(0;x)  & ~\sim~  \frac{1}{r^2} \,, \qquad p(-2;x)   ~\sim~  \frac{1}{r}  \,, \qquad p(-4;x)    ~\sim~  const \,, \qquad  p(-6;x)   ~\sim~ u^6 \, r^4  \,, \\ 
 p(a;x)   & ~\sim~ u^6 \, r^{1- \frac{a}{2}} \quad  \text{ for} \quad a \le -6 \,.
\end{aligned}
\label{pasymp}
\end{equation}
These functions are smooth and finite everywhere for $a \le -4$ and  vanish as $u,v \to \infty$ and as $ u, r \to 0$ for $a \le -6$.

Similarly, as $u \to 0$ (with $r$ finite) one has:
\begin{equation}
\begin{aligned}
p(a;x)  & ~\sim~   A_a \, u^6  \, r^{1- \frac{a}{2}}  \quad  \text{ for} \quad a \ge 8\quad  \text{ or} \quad a \le -6 \,,   \qquad   p(-4;x)  ~\sim~   A_4~+~ B_4 \, u^6  \, r^3   \\
 p(a;x)  & ~\sim~ A_a \ \, r^{-2- \frac{a}{2}} ~+~ B_a \ \, u^2 \, r^{-1- \frac{a}{2}} ~+~\cO(u^4)  \quad  \text{ for} \quad -2 \le a \le  6 \,,
\end{aligned}
\label{pasymp}
\end{equation} 
for some non-zero constants $A_a$ and $B_a$.  For $a=2$, one has
\begin{equation}
p(2; x)  ~=~ \frac{64\, \alpha^3} {r^3}    \,,
\label{peql2}
\end{equation}
which is $u$ independent.  One should note that this means that the polarization function, $k$, (see (\ref{kform2})) is smooth 
as $u \to 0$ for $a  \ge 8$ or  $a \le -4$, and diverges as $u^{-2}$ for $a=-2,0,4,6$ and vanishes identically for $a=2$.

We now normalize our Ansatz for $p$ by taking:
\begin{equation}
p(a;u, x)  ~=~ \frac{Q}{64\, \alpha^3} \, u^{a + 4} \,  F(a;x)   \,,
\label{pAnsatz2}
\end{equation}
where $F$ is given by (\ref{Fsol1}).

One can now substitute this into the source, $s$, defined by  (\ref{source1}) and we find:
\begin{equation}
\begin{aligned}
s  ~\equiv~ -\frac{Q^2\, u^{2(a-2)}}{128 } \,\frac{x^4}{(1+x)}  & \,\big[(a+4)\, F_1(x) ~-~  (a-2)\, F_2 (x) \big]\\
& \times \Big[(a+4)\,(1 + \coeff{1}{2}\, a\,x)\, F_1(x) ~-~  (a-2)\, (1 +  (a-1)\,x)\, F_2 (x) \Big] \,,
\end{aligned}
 \label{source2}
\end{equation}
where 
\begin{equation}
F_1(x)  ~\equiv~ {}_2F_1 \bigg(1- \frac{a}{2} , 4- \frac{a}{2}  , 4; -x \bigg) \,, \qquad F_2(x)  ~\equiv~ {}_2F_1 \bigg(2- \frac{a}{2} , 4- \frac{a}{2}  , 4; -x \bigg)  \,.
 \label{F1F2defns}
\end{equation}
One can then solve (\ref{Peqn}) for $P$ using the Ansatz: 
\begin{equation}
P  ~=~  Q^2\, u^{2(a-1)} \, K(x)   \,,
 \label{PAnsatz}
\end{equation}
which is consistent with the scaling behavior (\ref{scaling2}).  Indeed, one then finds
\begin{equation}
x(1+x) \, \frac{d^2 K}{dx^2}   ~-~2\,\big(1+(a-1)\, x\big) \, \frac{dK}{dx} ~+~a(a-1)  \, K ~=~ \sigma(x) \,,
\label{Keqn}
\end{equation}
where
\begin{equation}
\begin{aligned}
\sigma(x) ~\equiv~ 
 -\frac{1}{512} \,\frac{x^4}{(1+x)}  & \,\big[(a+4)\, F_1(x) ~-~  (a-2)\, F_2 (x) \big]\\
& \times \Big[(a+4)\,(1 + \coeff{1}{2}\, a\,x)\, F_1(x) ~-~  (a-2)\, (1 +  (a-1)\,x)\, F_2 (x) \Big] \,.
\end{aligned}
 \label{source3}
\end{equation}

For $a$ even, the source $\sigma(x)$ has the form $(1+x)^{-m} \, h(x)$, where $h(x)$ is a polynomial, and it is therefore relatively straightforward to solve (\ref{Keqn}) for any particular value of (even) $a$.  There is then the issue of how to select the homogeneous solutions.  Usually this is done in such a manner as to soften or remove singularities.

\subsection{The momentum-carrying flux}
\label{ss:Asymp}

To understand the behavior of the spherically symmetric, scaling solutions is is useful to catalog the  behavior of the flux function (see (\ref{C3part1})).  In general one has:
\begin{equation}
\cC ~=~   \frac{\big( \partial_z p \big) }  {4\, u^4  (-\partial_z w)}   ~=~   \frac{Q\, u^{a-4}  \cos \theta }{16 } \, \bigg( \frac{1}{16} \,(a-2)(a-8)\, x^2\, F_3(x) ~-~ 3\, x\, F_1(x)  \bigg) \,,
 \label{C3res3}
\end{equation}
where
\begin{equation}
F_3(x)   ~\equiv~ {}_2F_1 \bigg(2- \frac{a}{2} , 5- \frac{a}{2}  , 5; -x \bigg)  \,.
 \label{F3defn}
\end{equation}
%

\subsubsection{Specific solutions}
\label{ss:LocCsols} 

As we indicated earlier, we  restrict ourselves  to even integer values of $a$ since this leads to closed form expressions  for all the hypergeometric functions and produces a richer structure in the series expansions.  

For $a>0$, the expressions for $\cC$ are:
\begin{equation}
\begin{aligned}
a= 2:  \ \ &    - \frac{3\, Q\, \cos \theta }{4\, \alpha^3 \, u^4 \, r}  \,,\qquad\qquad\qquad   a = 4:   \ \    - \frac{Q\, (3\, r u^2 + 8\, \alpha^3)\cos \theta }{4\, u^4 \, r^2 } \,,  \\
a= 6:   \ \   & - \frac{ Q\, (3\, r^2 u^4+ 8\, \alpha^3 r u^2 + 8\, \alpha^6) \cos \theta }{4\, u^4 \, r^3 }  \,,\qquad\qquad a= 8:   \ \    - \frac{3\, Q\, u^2 \cos \theta }{4\, r} \,,  \\
a= 10:   \ \  & - \frac{Q\, u^2(3\, r u^2 - 16\, \alpha^3)\cos \theta }{4\,  r^2 }  \,, \quad   \ \  a= 12:   \    - \frac{Q\, u^2(3\, r^2 u^4 -40 \, \alpha^3 r u^2 + 80\, \alpha^6)\cos \theta }{4\,  r^3 }  \,,  \dots 
\end{aligned}
 \label{C3apos}
\end{equation}
while, for $a\le0$, the expressions for $\cC$ are:
\begin{equation}
\begin{aligned}
a= 0: \quad &  - \frac{Q\, (3\, r u^2 + 8\, \alpha^3)\cos \theta }{4\, u^4 \, ( r u^2 + 4\, \alpha^3)^2} \,,  \qquad
 a = -2: \quad  - \frac{ Q\, r\,(3\, r^2 u^4+ 8\, \alpha^3 r u^2 + 8\, \alpha^6) \cos \theta }{4\, u^4 \, ( r u^2 + 4\, \alpha^3)^4 }  \,,  \\
a= -4: \quad   &  - \frac{3\, Q\, u^2 \,r^5 \cos \theta }{4\, ( r u^2 + 4\, \alpha^3)^6 } \,, \qquad\qquad  \ \ 
a= -6: \quad   - \frac{Q\, r^6 \,u^2(3\, r u^2 - 16\, \alpha^3) \cos \theta }{4\, ( r u^2 + 4\, \alpha^3)^8} \,,  \\
a= -8: \quad & - \frac{Q\, r^7 u^2(3\, r^2 u^4 -40 \, \alpha^3 r u^2 + 80\, \alpha^6) \cos \theta }{4\, ( r u^2 + 4\, \alpha^3)^{10}}  \,,  \dots \,.
\end{aligned}
 \label{C3anonpos}
\end{equation}
%

\subsubsection{Localizing the flux function}
\label{ss:LocC} 

First and foremost, we note that for finite $u>0$,  all these solutions fall off  as $r^{-1}$ as $r \to \infty$, which means that the fluxes are concentrated around the M5 brane. 
Indeed, for $ a >0$, the solutions diverge as   $r \to 0$, which means that these fluxes have singular sources on the M5 brane.  On the other hand, for $a \le 0$, the solutions are smooth and finite for all values of $r$ (with $u >0$, fixed).  Indeed, the fluxes follow a ``bump-function'' in $r$,  whose profile and peak is determined by $u$ and $\alpha$.  These solutions appear to be exhibiting the  migration described in \cite{Bena:2025uyg}. 

The $u$-dependence also undergoes a major qualitative change when inside, or outside, the interval, $\cI$, defined by $-2 \le a \le 6$.  Inside this range, $\cC$ diverges as $u \to 0$, while outside this range, $\cC$ vanishes as $u \to 0$.   For $a > 6$, $\cC$ diverges as $u \to \infty$,  while, for $a \le 6$,  $\cC$ remains bounded as $u \to \infty$. 

We are, first and foremost, interested in modes that remain bounded for $u \to  \infty$  and so we take $a \le 6$.   For $a=2,4,6$, the fluxes have strongly localized (singular) sources at $r=0$ and $u=0$.  For $a\le -4$, the fluxes are localized by a bump function near  $r=0$ and $u=0$ but actually vanish at $u=0$ and at $r=0$.  For $a =0$, $\cC$ is  strongly localized (singular) at  $u=0$ and finite at $r=0$.   

There are also mixed scaling limits that take the bump functions for $a\le -2$ into strongly localized but singular limits.  Specifically, we can take $u \to 0$ and $r \to \infty$ and perhaps the most interesting way to do this is to keep $x$, or $r u^2$, fixed as  $u \to 0$ and $r \to \infty$.  This is geometrically interesting because it keeps the $S^3$,  defined by the $\sigma_j$ in  (\ref{11metric-AdS}), at a fixed scale.  As is evident from (\ref{C3res1}), in this limit, $\cC$ diverges as $u^{a-4}$ for $a <4$.
While this limit leads to fluxes that localize at $u=0$, this only happens as $r \to \infty$, and so the fluxes are localizing high in the AdS throat (large $r$).

The bottom line this that the fluxes are bounded at infinity so long as $a \le 6$.  For  $a \le 6$, $\cC$ is  strongly localized, with singular sources at $r=0$ and at $u=0$, for $a>0$.  For $a =0$, $\cC$ has a singular sources at $u=0$ and remains finite at $r=0$.  For $ a < 0$,   $\cC$ vanishes at $u=0$ and at $r=0$ but is localized by a bump function near the brane.  We will therefore focus on the solutions with $a \le 6$  (with, of course, $a \in 2\ZZ$).

We also recall that the polarization function, $k$, diverges as $u \to 0$ for $a=-4,-2,0,4,6$.  We not going to focus on this because it is caused by the collapse of the $S^3$ and is not a true signature of the localization of the excitations.

\section{Examples}
\label{sec:Examples}

We now exhibit a range of explicit examples that illustrate and extend the diverse  asymptotic behaviors that we started to catalog in Sections \ref{ss:simpsol2} and \ref{ss:Asymp}.

\subsection{The simplest solution}
\label{ss:example1}

The  simplest  example is to take $a=2$ and use 
\begin{equation}
p(u,v,z)  ~=~ \frac{Q}{64\, \alpha^9} \, u^{6} \, x^3  ~=~ \frac{Q  }{r^3}  ~=~ \frac{Q}{(v^2 +z^2)^{\frac{3}{2}}}   \,.
\label{pAnsatz3}
\end{equation}
This is harmonic on the $\IR^5$ transverse to the M5 brane:
\begin{equation}
 \frac{1}{r^4} \, \partial_r  \big(r^ 4\, \partial_r p \big)  ~+~\frac{1}{r^2 \, \sin^3 \theta } \, \partial_\theta \big(\sin^3 \theta \, \partial_\theta p \big) ~=~ 0 \,,
 \label{peqn-simp2}
\end{equation}
and it leads to the flux function (see (\ref{C3part1})): 
\begin{equation}
\cC ~=~   \frac{\big( \partial_z p \big) }  {4\, u^4  (-\partial_z w)}   ~=~  - \frac{3\, Q\, \cos \theta }{4\, \alpha^3 \, u^4 \, r} ~=~  - \frac{3\, Q\, x }{16\, \alpha^3 \, u^2} \,  \cos \theta \,.
 \label{C3part3}
\end{equation}
Also note that in this example, the polarization vector, $k$, vanishes identically (see, (\ref{kform2})), which means that this solution should represent some form of charge density wave.

The source, (\ref{source1}), becomes
\begin{equation}
s  ~=~  -\frac{72\, Q^2\, \alpha^6}{\,u^8 \, (v^2 +z^2)^{2} }\,,
\label{source-simp1}
\end{equation}
which has scaling dimension $0$.  
The general  solution of (\ref{Peqn}) is then:
\begin{equation}
P  ~=~-\frac{3\,Q^2\, u^2}{256\,\alpha^6} \,\Big[x^3(x -2 ) ~+~c_1 (1+x) ~+~c_2 \big(x^2 + x-  1 - 2(1+x)\log (1+x)\big)  \Big]\,,
\label{Ppart1}
\end{equation}
for some constants, $c_1$ and $c_2$. Setting these constants to zero leads to
\begin{equation}
P  ~=~\frac{3\,Q^2\,\alpha^3}{2} \, \bigg[  \frac{1}{2 u^4 \, (v^2 +z^2)^{\frac{3}{2} } } ~-~  \frac{2\,  \alpha^3}{u^6 \, (v^2 +z^2)^{2} } \bigg]  \,.
\label{Ppart2}
\end{equation}

This solution comes from a simple, singular harmonic source in $p$ (\ref{pAnsatz3}) and the flux function, $\cC$, and momentum function, $P$, are highly localized (singular) at $u=0$ and at $r=0$.  

The values of the $c_j$ do not affect the strength of the singularity at either $r=0$ or $u=0$, but one can choose the $c_j$ to obtain a more rapid fall off at infinity.  Indeed, taking $c_1 =c_2 =6$ leads to:
\begin{equation}
\begin{aligned}
P   ~\sim~  -\frac{36\, Q^2 }{5\, \alpha^{3} \, u^8 \, r^5 }  \,,
\end{aligned}
\label{Pasymp2}
\end{equation}
as $r \to \infty$ or  $u \to \infty$.

\subsection{Further solutions}
\label{ss:examples}

\subsubsection{Examples: $a = 4, 6$}
\label{ss:example2} 
For $a =4, 6$, the flux functions are: 
\begin{equation}
\cC ~=~   - \frac{Q \,(3\,   r\, u^2 + 8 \,\alpha^3)\,  \cos \theta  }{4\, r^2\, u^4 }  \,, \qquad \cC ~=~ - \frac{Q \,(3\,   r^2\, u^4 + 8 \,\alpha^3\, r\,u^2 + 8 \,\alpha^6)\,  \cos \theta  }{4\, r^3\, u^4 }\,.
 \label{C3part4}
\end{equation}
The momentum functions are rather more complicated.  For $a=4$ one has:
\begin{equation}
\begin{aligned}
P  = -\frac{Q^2\, u^6}{1536 \,\alpha^{12}} \,\Big[& -x( 156 +858  +1680 x^2 -1443 x^3 -36  x^4 + 16 x^5 ) \\
&+~156\,\Big((1+x)  (1 +5 x + 13 x^2-3 x^3 ) \, \log (1+x) + 12 x^3 \, {\rm Li}_2 (-x)  \Big) \Big] + P_{hom} \,,
\end{aligned}
\label{Ppart4}
\end{equation}
where Li$_2$ is the dilogarithm function and 
\begin{equation}
\begin{aligned}
P_{hom}   ~=~ -\frac{Q^2\, u^6}{1536 \,\alpha^{12}} \, \Big[\, c_1 ~+~ c_2 \, \big( (1 + 6x +18 x^2 -3 x^4) -12 x^3 \log x \big) \,  \Big]   \,.
\end{aligned}
\label{Phom4}
\end{equation}
Requiring that $P$ is bounded as $u \to \infty$ leads to $c_1 =c_2 =0$ and then 
\begin{equation}
\begin{aligned}
P   ~\sim~  \frac{89\, Q^2}{3 \,\alpha^{3} \, r^3 }  \,,
\end{aligned}
\label{Pasymp4}
\end{equation}
as $u \to \infty$, or as $r \to \infty$.   One then finds that 
\begin{equation}
\begin{aligned}
P   ~\sim~  -\frac{128\, Q^2 \, \alpha^{6}}{3 \, u^6 \, r^3 }  \,.
\end{aligned}
\label{Pasymp4a}
\end{equation}
as $u \to 0$, or as $r \to 0$.

For $a=6$ one has:
\begin{equation}
\begin{aligned}
P  = -\frac{Q^2\, u^{10}}{76\,800 \,\alpha^{12}} \,\Big[& x( 840  +12\,180x   +119\,980 x^2 -1\,319\,010 x^3 -35\,380  x^5 + 75 x^7 ) \\
&+840\,\Big((1+x)  (-1 -14 x -136 x^2 +1036 x^3 -511 x^4 +10 x^5 ) \, \log (1+x)\\
&  + 50\,400 \, x^3 \,(10 -15 x +3 x^2 )\, {\rm Li}_2 (-x)  \Big) \Big] + P_{hom} \,,
\end{aligned}
\label{Ppart6}
\end{equation}
and 
\begin{equation}
\begin{aligned}
P_{hom}   ~=~ -\frac{Q^2\, u^6}{1536 \,\alpha^{12}} \, \Big[&\, c_1 \,x^3 \,(10 -15 x +3 x^2 )\,  \\
&+~ c_2 \, \Big( (-3 -45 x -450 x^2 +7210 x^3 -5190 x^4 -150 x^5 + 30 x^6) \\
&  \qquad\qquad +180 x^3\,(10 -15 x +3 x^2 )\, \log x\, \Big) \,  \Big]   \,.
\end{aligned}
\label{Phom6}
\end{equation}
As before, requiring that $P$ is bounded as $u \to \infty$ leads to $c_1 =c_2 =0$ and then one has
\begin{equation}
\begin{aligned}
P   ~\sim~  -\frac{36\, Q^2\,\alpha^{3}}{5  \, r^5 }  \,.
\end{aligned}
\label{Pasymp6}
\end{equation}
as $u \to \infty$, or as $r \to \infty$.   For $u \to 0$, or as $r \to 0$, one then finds: 
\begin{equation}
\begin{aligned}
P   ~\sim~  -\frac{64\, Q^2 \, \alpha^{12}}{3 \, u^6 \, r^8 }  \,.
\end{aligned}
\label{Pasymp6a}
\end{equation}
%

\subsubsection{Examples: $a = 0, -2, -4$}
\label{ss:example3} 

For $a =0, -2$, we  find flux functions fall off as  $u,r \to \infty$ but still diverge as $u^{-4}$ as $u \to 0$: 
\begin{equation}
\cC ~=~   - \frac{Q\,  \cos \theta \,(3\,   r\, u^2 + 8 \,\alpha^3) }{4\, u^4 \, (r\, u^2 + 4\alpha^3)^3  }\,, \qquad \cC ~=~   - \frac{Q\, r\,  \cos \theta \,(3\, r^2 u^4 + 8\, \alpha^3 r\, u^2 +  8 \,\alpha^6 ) }{4\, u^4 \, (r\, u^2 + 4\,\alpha^3 )^4  } \,,
 \label{C3part0-2}
\end{equation}
respectively. However, for $a=-4$, the flux function is  completely smooth, vanishing as $u, r \to 0$ and  $u,r \to \infty$:
\begin{equation}
\cC ~=~   - \frac{3\, Q\,u^2 r^5 \,  \cos \theta  }{4\, (r\, u^2 + 4\,\alpha^3 )^6  } \,.
 \label{C3part-4}
\end{equation}

The momentum functions are rather more complicated.  For $a=0$ one has:
\begin{equation}
\begin{aligned}
P  ~=~ \frac{Q^2}{128\,\alpha^{12}\, u^2} \,\bigg[& \frac{3 x( 4 + 14 x + x^2 - 10 x^3 - 2 x^4 ) }{ 6\,(1+x)^4 }\\
&~-~ \frac{2 +6 x +6 x^2 + x^3  }{ (1+x)^3}\, \log (1+x) \bigg] ~+~ P_{hom} \,,
\end{aligned}
\label{Ppart3}
\end{equation}
where 
\begin{equation}
P_{hom}   ~=~ \frac{Q^2}{128\,\alpha^{12}\, u^2} \,\bigg[\, c_1 ~+~ c_2 \, \frac{ x^3}{(1+x)^3 }  \,  \bigg]   \,,
\label{Ppart3a}
\end{equation}
for any constants $c_1$ and $c_2$.

To ensure that $P$ vanishes as $r \to \infty$ one must take $c_1=0$, and then $P \sim r^{-4} u^{-10}$.  However, taking $c_2 = \frac{49}{6}$ leads to an even stronger fall-off:  
\begin{equation}
P  ~\sim~ -\frac{36\,\alpha^{3}\, Q^2}{5\, u^{12}\, r^5}  \,,
\label{Pasymp01}
\end{equation}
as either $u \to \infty$ or $r \to \infty$.

 Independent of  $c_1$ and  $c_2$, as either $u \to 0 $ or $r \to 0$,  one finds:
\begin{equation}
P  ~\sim~ -\frac{Q^2}{32\,\alpha^{9}\, u^4\, r}  \,,
\label{Psrc0}
\end{equation}
which is a softer singularity than that of (\ref{Ppart1}).

For $a=-2$ one has:
\begin{equation}
\begin{aligned}
P  =- \frac{Q^2\, x^3}{1792\, \alpha^{12}\, u^6} \,\bigg[& \frac{30\,294 +15\,267 x -14\, 391x^2 - 554 x^3 - 280 x^4 - 35  x^5 }{ 60\,(1+x)^8 }\\
& ~+~ \frac{(x-2)   }{  (1+x)^7}\, \log (1+x ) \bigg] ~+~ P_{hom} \,,
\end{aligned}
\label{Ppart-2}
\end{equation}
where 
\begin{equation}
\begin{aligned}
P_{hom}   =   & \frac{Q^2}{1792\,\alpha^{12}\, u^6} \,\bigg[\,  c_1 ~+~ c_2 \, \frac{\big(1 +10 x   +60 x^2 - 306 x^3 - 12 x^4 + 6  x^5 + 60\,x^3 \, (x-2) \log x  \big)}{  (1+x)^7}  \,  \bigg]   \,,
\end{aligned}
\label{Phom-2}
\end{equation}
for any constants $c_1$ and $c_2$.  

One also finds that, independent of $c_1$ and $c_2$, as $r \to 0$ or $u \to 0$, one has:
\begin{equation}
P  ~\sim~ \frac{Q^2}{3072 \,\alpha^{12}\, u^6}  \,.
\label{Psrc-2}
\end{equation}

To ensure that $P$ vanishes as $r \to \infty$ one must take $c_2=0$, and then $P \sim r^{-3} u^{-12}$.  However, taking $c_1 = -\frac{5049}{10}$ leads to an even stronger fall-off:  
\begin{equation}
P  ~\sim~ \frac{15\, 147 \, Q^2}{28\, u^{14}\, r^4}  \,,
\label{Pasymp-2}
\end{equation}
as either $u \to \infty$ or $r \to \infty$.

For $a=-4$ one has:
\begin{equation}
\begin{aligned}
P  ~=~ \frac{Q^2\,x^3}{14\,080\,\alpha^{12}\, u^{10}} \,\bigg[&  \frac{-6095  +12\, 610  x +3492 x^2 -1380 x^3 - 24 x^4+3 x^5  }{(1+x)^{12} }\\
& ~+~ \frac{84\, \,(-5 +15x  - 9 x^2 + x^3) }{  (1+x)^{11}}\, \log (1+ x) \bigg] ~+~ P_{hom} \,,
\end{aligned}
\label{Ppart-4a}
\end{equation}
where 
\begin{equation}
\begin{aligned}
P_{hom}   = \frac{Q^2}{\alpha^{12}\, u^6\,  (1+x)^{11}} \,\bigg[\,   & c_1\, x^3 \,(-5 +15x  - 9 x^2 + x^3)  \\
&  +c_2 \, \Big((1 +21 x +315 x^2 - 10\,595 x^3 - 20\, 235 x^4 - 6\, 345  x^5 \\&  - 135  x^6 +  15 x^7) + ~420 \,x^3 \, \,(-5 +15x  - 9 x^2 + x^3)\, \log x)  \Big)  \,  \bigg]   \,,
\end{aligned}
\label{Phom-4}
\end{equation}
for any constants $c_1$ and $c_2$.

The solution is finite  as $r \to \infty$ and falls off as $u^{-10}$ as  $u \to \infty$, and,  by judicious choice of $c_1$ and $c_2$, can be made to fall off as $r^{-3} u^{-16}$  as $r \to \infty$.  However, $c_1$ and $c_2$ now enter into the behavior as $u \to 0$ or $r \to 0$.  Indeed, taking  $c_1 =11, c_2 = -\frac{1}{5}$  one finds that
as either $u \to 0$ or $r \to 0$ one has
\begin{equation}
P  ~\sim~ \frac{Q^2\, r^5 }{36\,700\,160 \,\alpha^{27}}  \,,
\label{Psrc-4}
\end{equation}
which is completely smooth.  Explicitly, for  $c_1 =11, c_2 = -\frac{1}{5}$, one has 
\begin{equation}
\begin{aligned}
P  =- \frac{Q^2}{1\,971\,200 \,\alpha^{12}\, u^{10}} \,\bigg[&  \frac{ 1+ 22  x +336 x^2+ 20\, 470 x^3 - 53\, 960 x^4 - 3900 x^5+860 x^6  -55 x^7  }{(1+x)^{12} }\\
& ~+~ \frac{420\, x^3\, (5 -15x  + 9 x^2 - x^3) }{  (1+x)^{11}}\, \log \bigg(1+ \frac{1}{x}\bigg) \bigg]\,,
\end{aligned}
\label{Ppart-4b}
\end{equation}
and at  large $r$ or large $u$, the leading term is:
\begin{equation}
P  ~\sim~- \frac{Q^2}{1\,971\,200 \,\alpha^{12}}\,\frac{r^{12} \, u^{14}}{(r \, u^2 + 4 \, \alpha^3)^{12}} \,,
\label{Pasymp-4}
\end{equation}
which is completely smooth.

Thus the solution flux function, $\cC$, (see (\ref{C3part-4})) and momentum function, $P$,  are completely smooth ``bump functions'' for $a=-4$.   

\subsubsection{Examples: $a = -6$}
\label{ss:example4} 

The flux function is again completely smooth and is now vanishing as $u, r \to 0$ and  $u,r \to \infty$:
\begin{equation}
\cC ~=~   - \frac{Q\,u^2 \,r^6 \,  (3 u^2 \,r  -16 \alpha^3) \,\cos \theta  }{4\, (r\, u^2 + 4\,\alpha^3 )^8  }~=~   - \frac{Q}{16\, \alpha^3 \, u^{10}}\,\frac{x(3 -4 x)  }{(1+x)^8  }  \,\cos \theta  \,.
 \label{C3part5}
\end{equation}

As with $a=-4$ , the solution  is finite as $r \to \infty$ and falls off fast as $u \to \infty$, independent of the choice of homogeneous solutions.  There are choices of the homogeneous solution that lead to faster fall off at infinity, however, the generic solution is singular at $u=0$.  One can choose  the homogeneous solution to remove this singularity, and make the solution vanish at  $u \to 0$ and at $r \to 0$.  The resulting solution is:
\begin{equation}
\begin{aligned}
P  ~=~- & \frac{Q^2}{38\,438\,400\, \alpha^{12}\, u^{14}\,(1+x)^{16} }  \\
& \qquad\times \bigg[ \big(5 +185 x +5220 x^2 + 136\,752 x^3 - 1\,782\, 228 x^4 \\
&\qquad \qquad+  3\, 824 \, 730 x^5 - 1\, 773 \, 660 x^6 +147\, 420 x^7+ 7560 x^8 \big) \\
& \qquad \quad\ \ ~+~ 2520 \, x^3\,(1+x) \,(28 -210x+ 420  x^2 -280 x^3+ 60 x^3 -3 x^5)\, \log \bigg(1+\frac{1}{x}\bigg) \bigg] \,.
\end{aligned}
\label{Ppart-6}
\end{equation}
 At  large $r$, or large $u$, the leading term in this solution is:
\begin{equation}
P  ~=~ - \frac{Q^2\, u^{18}\, r^{16} }{7\,687\,680 \,\alpha^{12} \, (r \, u^2 + 4 \, \alpha^3)^{16} } ~+~ \dots \,,
\label{Pasymp-6}
\end{equation}
 At   $r \to 0$, or  $u \to 0$, one has:
\begin{equation}
P  ~\sim~ - \frac{Q^2\, u^{6}\, r^{10} }{16\,777\,216 \,\alpha^{42}}\,.
\label{Psrc-6}
\end{equation}
Thus $\cC$ and $P$ are  smooth.

\subsubsection{A comment on the polarization function}
\label{ss:polfn} 

It is interesting to recall what we observed about the polarization function, $k$, in the discussion around (\ref{pasymp}):  This function is smooth 
as $u \to 0$ for $a  \ge 8$ or  $a \le -4$, and diverges as $u^{-2}$ for $a=-2,0,4,6$ and vanishes identically for $a=2$.    
We see the underlying reason for this in the behavior of the flux function, $\cC$, and the momentum density, $P$, as $u \to 0$.  When these functions are singular as $u \to 0$ then so is the polarization function, and when $\cC$ and $P$ are finite as $u \to 0$, then $k$ vanishes in this limit.  Indeed, $\cC$ is either singular or vanishing as $u \to 0$ and this exactly correlates with the behavior of $k$.    

Simply put, when the fluxes are localized via a singularity in $u$ then the polarization vector is similarly singular.  However, when when the flux function is localized away from $u=0$ through a bump function, then $k$ is smooth and vanishing at $u=0$.

It is also worth noting that as $u \to \infty$ one has:
\begin{equation}
k  ~\sim~ (a-2) \,  \frac{Q\,\alpha^{3}  }{ 2\,r^3} \, u^{a-6}\,.
\label{kinf1}
\end{equation}
Just like the flux function, $\cC$, this  diverges as $u \to \infty$ for  $a > 6$ and remains bounded for $a \le 6$. Thus reinforcing our restriction of solutions with $a \le 6$.

\subsection{One last example: breaking spherical symmetry}
\label{ss:example5} 

One can explore higher harmonics, for example, taking:
\begin{equation}
 p(u,v, z)  ~=~ \frac{Q\,\cos \theta }{r^4}  ~=~ \frac{Q \, z}{(v^2 +z^2)^{\frac{5}{2} } }\,,
 \label{p-simp3}
\end{equation}
This also leads to a simple flux function: 
\begin{equation}
\cC ~=~   \frac{\big( \partial_z p \big) }  {4\, u^4  (-\partial_z w)}   ~=~  \frac{Q\,(v^2-4 z^2)}{4\, \alpha^3 \, u^4 \, (v^2 +z^2)^2} \,,
 \label{C3part2}
\end{equation}
and source for $P$  is given by:
\begin{equation}
s  ~=~    -\frac{32\, Q^2\, (v^2 +16 z^2)}{\alpha^6\,u^8 \, (v^2 +z^2)^{4} } ~=~    -\frac{32\, Q^2\, (1 +15 \cos^2 \theta )}{\alpha^6\,u^8 \, r^{6} } ~=~    -\frac{Q^2\,u^4  \, x^6  \, (1 +15 \cos^2 \theta )}{128\,  \alpha^{24} } \,, 
\label{source-simp2}
\end{equation}
 which has scaling dimension, $-4$.  
 
 To solve for $P$ one simply uses  an Ansatz with and appropriate power of $u$, the correct spherical harmonics and two arbitrary functions of the scale-invariant variable, $x$:   $P = u^6 (H_1(x) +  H_2(x) \cos^2 \theta)$.
\begin{equation}
\begin{aligned}
P  ~=~ - \frac{Q^2 \, u^6}{256\,\alpha^{24} } \, \bigg[& \frac{5\, x^{5}}{4} \, (x-4) \, \cos^2 \theta \\
& +    \frac{1}{12}\, (1+x) \,  x\,\big( 120 +540 x -1085 x^2- 25 x^3 +x^4\big) \\
&- 10\, (1+x) \,\big(1+5x +13 x^2-3x^3\big)\, \log(1+x) ~-~ 120 \,x^3 \, {\rm Li}_2(-x)\ \bigg]~+~P_{hom}  \,.
\end{aligned}
\label{Pnonsymm1}
\end{equation}
The homogeneous solution now has two parts with different harmonic pieces:
\begin{equation}
\begin{aligned}
P_{hom}  ~=~ - \frac{Q^2 \, u^6}{256\,\alpha^{24} }& \, \bigg[ c_1 \, x^3 ~+~ c_2\, \big(1 + 6 x + 18 x^2 - 3  x^4 -12 x^3 \log x \big) \\
&+\frac{(1 - 5\, \cos^2 \theta)}{x^2} \, \Big( c_3\, (1+x)^5  +~ c_4 \,\big( 87 +375 x + 600 x^2 + 400 x^3 + 50 x^4 \\
& \qquad\qquad\qquad\qquad\qquad\qquad -50x^5 -10 x^6 + 60\, (1+x)^5 \log(1+x)  \big)\Big)\bigg]   \,.
\end{aligned}
\label{Phomnonsymm1}
\end{equation}
The singularity as either $r\to 0$ or $u \to 0$ is independent of the $c_j$ and has the form:
\begin{equation}
P  ~\sim~ - \frac{4\, Q^2  }{ 3\,\alpha^{6}\, u^{6}\, r^{6} }\, (1 + 15\, \cos^2 \theta) \,.
\label{Psrc-nonsymm}
\end{equation}
As before one can choose the $c_j$ so as to generate the strongest fall off as $r \to \infty$ and $u \to \infty$.  Indeed, taking $c_1 =\frac{795}{4}\,, c_2 =0 \,, c_3 =-\frac{609}{10}$ and $c_4 =\frac{7}{10}$, one arrives at: 
\begin{equation}
P  ~\sim~  \frac{16\, Q^2  }{ 21\,\alpha^{3}\, u^{8}\, r^{7} }\, (1 - 35\, \cos^2 \theta) \,.
\label{Pasymp-nonsymm}
\end{equation}
as either $r \to \infty$ or $u \to \infty$.

The significance of this example is that it suggests that, for any value of $a$, there are solutions for general harmonics on the $S^4$ and that, while they are  more complicated,  they have very similar properties to spherically symmetric, ``S-wave'' solutions discussed in the first parts of this section.

%
\section{Final comments}
\label{sec:Conclusions}

Our purpose here has been to explore the rather abstract momentum waves of the \nBPS{8}  M2-M5-P  system discovered in \cite{Bena:2024qed}, but in a much simpler setting in which one can construct a range of examples. We have shown the even in simplest possible non-trivial substrate, the near-brane limit of M5 branes, the linear system of \cite{Bena:2024qed} captures the symmetric null wave solutions that possess the M2-M5-P supersymmetries  defined by (\ref{projs3}).  We have also exhibited a large range of physically interesting solutions.  We chose the  M5 substrate not only because of its simplicity but because M5 branes locally dominate the substrate solutions analyzed in \cite{Bena:2023rzm,Bena:2024qed,Bena:2024dre,Bena:2025hxt} and so it makes the near-M5 brane the natural starting point to study localized momentum waves on M2-M5 intersections. 

These M5-P solutions, and the harmonic analysis,  are naturally organized by  sectioning the AdS$_7 \times S^4$  into the AdS$_3 \times \IR^4 \times \IR^4$  (see, for example,  (\ref{11metric-subs-simp1})) inherited from the generic M2-M5 system and then using the scale invariance of the substrate.  It is evident from the results presented here that there is a large collection of  solutions exhibiting quite a range of possible asymptotic behaviors.  In particular, we have shown how the momentum waves  not only localize on the M5 branes ($r \to 0$) but also localize on the AdS$_3$ factor ($u \to 0$).  We have also seen how this localization comes in two forms: (i) singular sources at  $r= 0$ and  $u = 0$, and (ii)  smooth bump functions that localize near $r,u= 0$.   The latter behavior appears to be another instance of momentum migration \cite{Bena:2025uyg}.  We also find solutions that have smooth momentum and flux distributions away from the brane sources.

We have studied this rather simple family of examples to see if the proposal of \cite{Bena:2024qed} can be  realized  in practice, and given our results, it is now important to test and generalize the momentum-wave proposal in new, non-trivial ways.  First and foremost, we have focussed on Choice (ii)  in \cite{Bena:2024qed}, in which the polarization vector $k$, is expected to be singular as $u \to 0$. Remarkably, we found that when the momentum wave migrates into a bump function, the polarization vector is actually smooth.  More generally, there is Choice (i) of \cite{Bena:2024qed}, but the near-brane limit of this choice is far from simple.  There are also a huge range of \nBPS{4}  brane intersections, and  near-brane limits, that can be used as substrates for \nBPS{8} momentum wave excitations.  There is a vast literature on such \nBPS{4}  brane intersections, and there is a forthcoming catalog \cite{BDTW} of such brane systems and near-horizon limits that can be related to the M2-M5-M5' system. It would be very interesting to generalize and study the other dual avatars of the M2-M5-M5'-P system.

Another aspect of the microstate geometry program is to understand how smooth geometries can emerge from \nBPS{8} momentum wave excitations of \nBPS{4}  substrates.  Here we have focussed on singular geometries because, despite their singularities, they are {\it locally} \nBPS{2} solutions, whose microstructure can be understood in terms of themelia \cite{Bena:2022fzf, Bena:2024qed}.  In practical terms, this means we have been constructing null waves on Poincar\'e AdS backgrounds.  It would be extremely interesting to see if one can construct momentum waves in {\it global} AdS$_3$, generalizing the superstratum construction to the M2-M5-M5'-P system.  Such a construction could have very interesting implications for the holography of the large $\cN=4$ CFT.  

Even more broadly, the \nBPS{8} M2-M5-M5'-P system  can be given a vastly richer structure,  involving further non-trivial geometry and fluxes,  without breaking supersymmetry.  Put differently, there is a broad range of \nBPS{8} substrates to which one can add a momentum wave without further breaking the supersymmetry.  Indeed, the projectors (\ref{projs3}) and  (\ref{projs2})  are compatible with 
\begin{equation}
 \Gamma^{2} \, \varepsilon  ~=~  \varepsilon \,,   \qquad  \Gamma^{3567} \, \varepsilon  ~=~- \varepsilon \,,\qquad \Gamma^{489\,10} \, \varepsilon  ~=~-  \varepsilon \,.
 \label{projs4}
\end{equation}
The first of these corresponds to a possible domain wall, while the last two suggest that the $\IR^4$'s can be replaced by hyper-K\"ahler spaces and that one can introduce self-dual fluxes on those spaces in a manner highly reminiscent of five-dimensional microstate geometries.  Such self-dual fluxes might also play an important role in carrying momentum.

It is evident that there are many options for developing \nBPS{8} momentum-carrying themelia and microstate geometries.  The art is going to be to find the generalizations that tell us the most about microstate structure and, or, states in holographic CFT's.

\vspace{1em}\noindent {\bf Acknowledgements:} 
 This work was supported in part by the ERC Grant 787320 - QBH Structure and by the DOE grant DE-SC0011687.  


\begin{adjustwidth}{-1mm}{-1mm} 

\bibliographystyle{utphys}      

\bibliography{references}

\end{adjustwidth}


\end{document}